\documentclass{ws-procs975x65}

\usepackage{xspace}
\newcommand{\FlukaLong}{{\scshape Fluka2008}\xspace}
\newcommand{\UrqmdLong}{{\scshape Urqmd1.3.1}\xspace}
\newcommand{\VenusLong}{{\scshape Venus4.12}\xspace}

\begin{document}

\title{Measurements of Cross Sections and Charged Pion Spectra in Proton-Carbon Interactions at 31 GeV/c}

\author{L. S. Esposito$^*$, on behalf of the NA61/SHINE Collaboration}

\address{ETH, Institute for Particle Physics,\\
Zurich, Switzerland\\
$^*$E-mail: luillo@cern.ch}


\author{The NA61 Collaboration:\\
N.~Abgrall${}^{23}$,
A.~Aduszkiewicz${}^{4}$,
B.~Andrieu${}^{12}$,
T.~Anticic${}^{14}$,
N.~Antoniou${}^{19}$,
J.~Argyriades${}^{23}$,
A.~G.~Asryan${}^{16}$,
B.~Baatar${}^{9}$,
A.~Blondel${}^{23}$,
J.~Blumer${}^{11}$
% ...
}

\author{
${}^{1}$ETH, Zurich, Switzerland \\
${}^{2}$Fachhochschule Frankfurt, Frankfurt, Germany \\
${}^{3}$Faculty of Physics, University of Sofia, Sofia, Bulgaria \\
${}^{4}$Faculty of Physics, University of Warsaw, Warsaw, Poland \\
${}^{5}$High Energy Accelerator Research Organization (KEK), Tsukuba, Ibaraki 305-0801, Japan \\
${}^{6}$Institute for Nuclear Research, Moscow, Russia \\
${}^{7}$Jagiellonian University, Cracow, Poland \\
${}^{8}$Jan Kochanowski University in  Kielce, Poland \\
${}^{9}$Joint Institute for Nuclear Research, Dubna, Russia \\
${}^{10}$KFKI Research Institute for Particle and Nuclear Physics, Budapest, Hungary \\
${}^{11}$Karlsruhe Institute of Technology, Karlsruhe, Germany \\
${}^{12}$LPNHE, University of Paris VI and VII, Paris, France \\
${}^{13}$Pusan National University, Pusan, Republic of Korea \\
${}^{14}$Rudjer Boskovic Institute, Zagreb, Croatia \\
${}^{15}$Soltan Institute for Nuclear Studies, Warsaw, Poland \\
${}^{16}$St. Petersburg State University, St. Petersburg, Russia \\
${}^{17}$State University of New York, Stony Brook, USA \\
${}^{18}$The Universidad Tecnica Federico Santa Maria, Valparaiso, Chile \\
${}^{19}$University of Athens, Athens, Greece \\
${}^{20}$University of Bergen, Bergen, Norway \\
${}^{21}$University of Bern, Bern, Switzerland \\
${}^{22}$University of Frankfurt, Frankfurt, Germany \\
${}^{23}$University of Geneva, Geneva, Switzerland \\
${}^{24}$University of Silesia, Katowice, Poland \\
${}^{25}$Warsaw University of Technology, Warsaw, Poland \\
${}^{a}${Present affiliation: University of Wroc{\l}aw, Wroc{\l}aw, Poland}\\
${}^{b}${\it deceased}\\
}


\begin{abstract}
As neutrino long baseline experiments enter a new domain of precision, the careful study of systematic errors due to poor knowledge of production cross sections for pions and kaons require more dedicated measurements for precise neutrino flux predictions.
The cosmic ray experiments require dedicated hadron production measurements to tune simulation models used to describe air shower profiles.
Among other goals, the NA61/SHINE (SPS Heavy Ion and Neutrino Experiment) experiment at the CERN SPS aims at precision measurements (5\% and below) for
both neutrino and cosmic ray experiments: those will improve the prediction of the neutrino
flux for the T2K experiment at J-PARC and the prediction of muon production in the
propagation of air showers for the Auger and KASCADE experiments.
NA61/SHINE took data during a pilot run in 2007 and in 2009 and 2010 with different carbon targets. 
The NA61/SHINE set-up and spectra for positive and negative pions obtained with the 2007 thin (4\% interaction length) carbon target data are presented~\cite{ref:pion_paper}.
\end{abstract}

\keywords{p+C interaction, inelastic cross section, inclusive pion spectra}

\bodymatter

\section{Physics motivation}
The NA61/SHINE (SPS Heavy Ion and Neutrino Experiment) experiment
 at the CERN SPS pursues a rich physics program in various
 fields~\cite{ref:proposala,ref:add1,ref:proposalb,ref:Status_Report_2008}.
 First, precise hadron production measurements are performed for
 improving calculations of the neutrino flux in the T2K neutrino
 oscillation experiment~\cite{ref:T2K},  as well as
 for more reliable simulations of cosmic-ray air showers in the Pierre Auger and KASCADE
 experiments~\cite{ref:Auger,ref:KASCADE}.
 Second, p+p, p+Pb and nucleus+nucleus collisions will be studied extensively
 at SPS energies. 
This article presents first NA61/SHINE results on charged
 pion spectra in p+C interactions at 31~GeV/c which are needed
 for an accurate neutrino flux prediction in the T2K experiment.
 The results are based on the data collected during the first NA61/SHINE
 run in 2007.

 T2K is a long baseline neutrino experiment in Japan, which uses
 a high intensity neutrino beam produced at J-PARC\footnote{Japan Proton
 Accelerator Research Complex 
 organized 
 jointly by JAEA and KEK in Tokai, Japan}.
 It
 aims to precisely measure the $\nu_\mu \to \nu_e$ appearance
 and $\nu_\mu$ disappearance~\cite{ref:T2K}.
 In order to generate the neutrino beam a high intensity
 30~GeV (kinetic energy) proton beam impinging on a 90~cm long graphite target is used,
 where $\pi$ and K mesons decaying into (anti)neutrinos are produced.
 The neutrino fluxes and spectra are then measured both at the near detector
 complex, 280~m from the target, and by the Super-Kamiokande (SK) detector
 located 295~km away from the neutrino source and 2.5 degrees off-axis.
 Neutrino oscillations are probed by comparing the neutrino flux
 measured at SK to the predicted one. 
 In order to predict the flux at SK one uses the near detector measurements
 and extrapolates them to SK with the help of Monte Carlo  simulations.
 Up to now, these Monte Carlo predictions
 are based on hadron production models only.
 For more precise predictions,
 which would allow the reduction of systematic uncertainties to the level needed
 for the T2K physics goals,
 measurements of pion and kaon production off carbon targets 
 are essential~\cite{ref:proposala,ref:add1,ref:proposalb}.
 The purpose of the NA61/SHINE measurements for T2K is to provide 
 this information
 at exactly the proton extraction energy of the J-PARC Main Ring
 synchrotron, namely 30~GeV kinetic energy (approximately 31~GeV/c momentum).
 Presently, the T2K neutrino beam-line is set up to focus positively charged
 hadrons, in such a way that it produces a $\nu_\mu$ beam.
 Spectra of positively charged pions presented in this paper constitute
 directly an essential ingredient in the neutrino flux calculation.
 
\section{The NA61/SHINE detector}
\begin{figure*}[!ht]
\begin{center}
\includegraphics[width=0.9\textwidth]{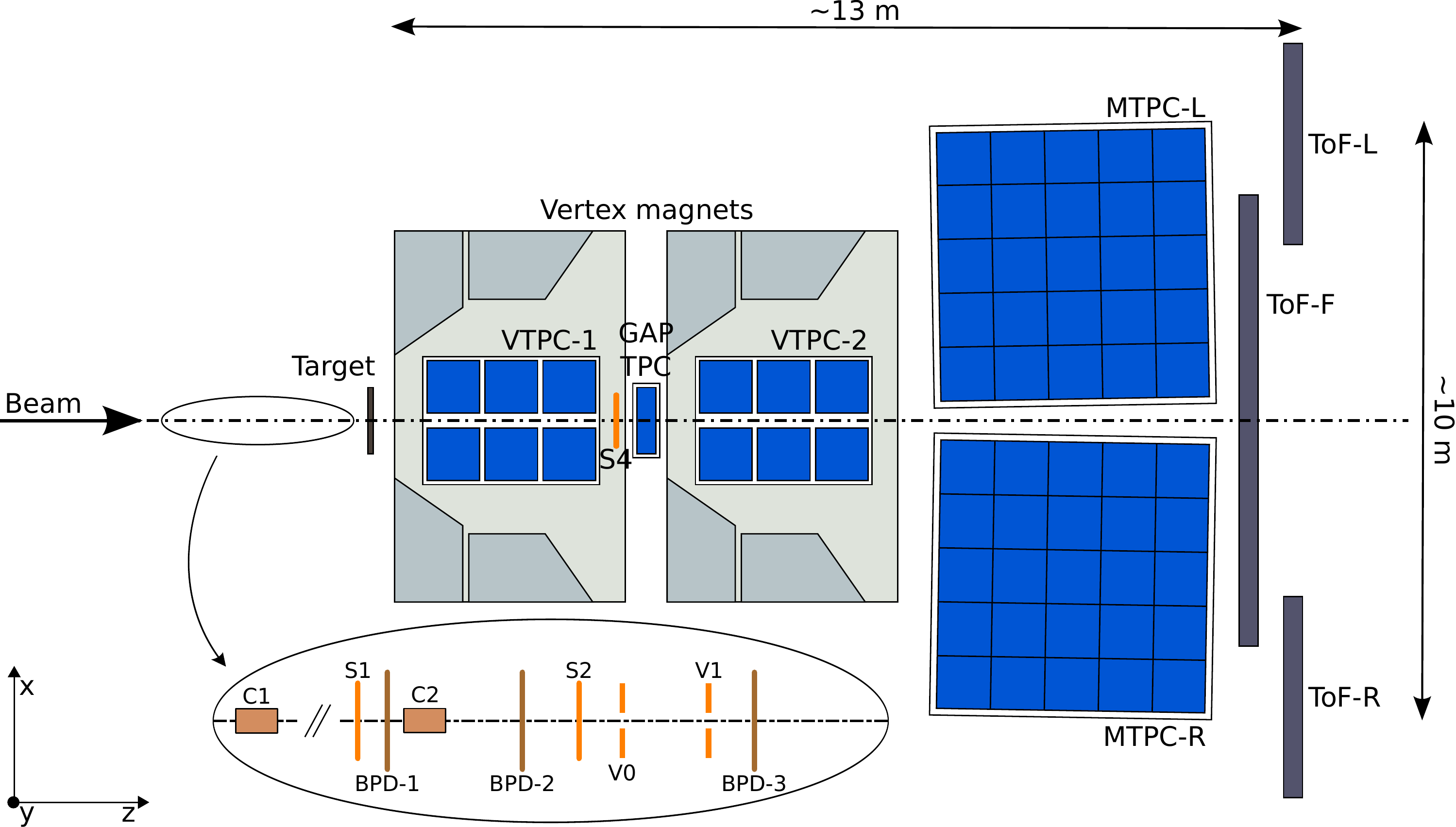}
\end{center}
  \caption{(Color online)
The layout of the NA61/SHINE experiment at the CERN SPS (top view, not to scale).
The chosen right-handed coordinate system is shown on the plot.
The incoming beam direction is along the $z$ axis.
The magnetic field bends charged particle trajectories
in the $x-z$ (horizontal) plane.
The drift direction in the TPCs is along the $y$ (vertical) axis.
}
  \label{fig:detector}
\end{figure*}

The NA61/SHINE experiment is a large acceptance hadron spectrometer in
the North Area H2
beam-line of the CERN SPS.
The schematic layout is shown in Fig.~\ref{fig:detector} 
together with the overall dimensions.

The main components of the current detector were constructed and used by the
NA49 collaboration~\cite{ref:NA49}.
A set of scintillation and Cherenkov counters as well as beam position
detectors (BPDs) upstream of the spectrometer provide timing reference,
identification and position measurements of the incoming beam particles. 
The main tracking devices of the spectrometer
are large volume Time Projection Chambers (TPCs).
Two of them, the vertex TPCs (VTPC-1 and VTPC-2 in Fig.~\ref{fig:detector}),
are located in
a free gap of 100~cm between the upper and lower coils of the
two superconducting dipole magnets.
Their maximum combined bending power is 9~Tm.
In order to optimize the acceptance of the detector at 31 GeV/c
beam momentum, the magnetic field used during the 2007 data taking period
was set to a bending power of 1.14~Tm. 
Two large TPCs (MTPC-L and \mbox{MTPC-R}) are positioned downstream of
the magnets symmetrically to the beam line. The
TPCs are filled with  Ar:CO$_2$ gas mixtures in proportions 90:10 for VTPCs
and 95:5 for MTPCs.
The particle identification capability of the TPCs
based on measurements of the specific energy loss, $dE/dx$,
is augmented by time-of-flight measurements using
Time-of-Flight (ToF) detectors.
The ToF-L and \mbox{ToF-R} arrays of scintillator pixels
have a time resolution of better than 90~ps~\cite{ref:NA49}.
Before the 2007 run the experiment was upgraded with a new
forward time-of-flight detector (ToF-F) in order to extend the acceptance.
The ToF-F consists of 64 scintillator bars with 
photomultiplier (PMT) readout at both ends 
resulting in a time resolution of about 115~ps.
The target under study is installed 80~cm in front of the VTPC-1.
The results presented here were obtained with an
isotropic graphite target of
dimensions 2.5(W)$\times$2.5(H)$\times$2(L)~cm and
with a density of $\rho = 1.84$~g/cm$^3$.
The target thickness along the beam is equivalent to about 4\% of
a nuclear interaction length~($\lambda_{\mathrm{I}}$).

\section{Analysis techniques}
\label{sec:analysis}

\begin{figure}[h!]
\begin{center}
\includegraphics[width=0.45\linewidth]{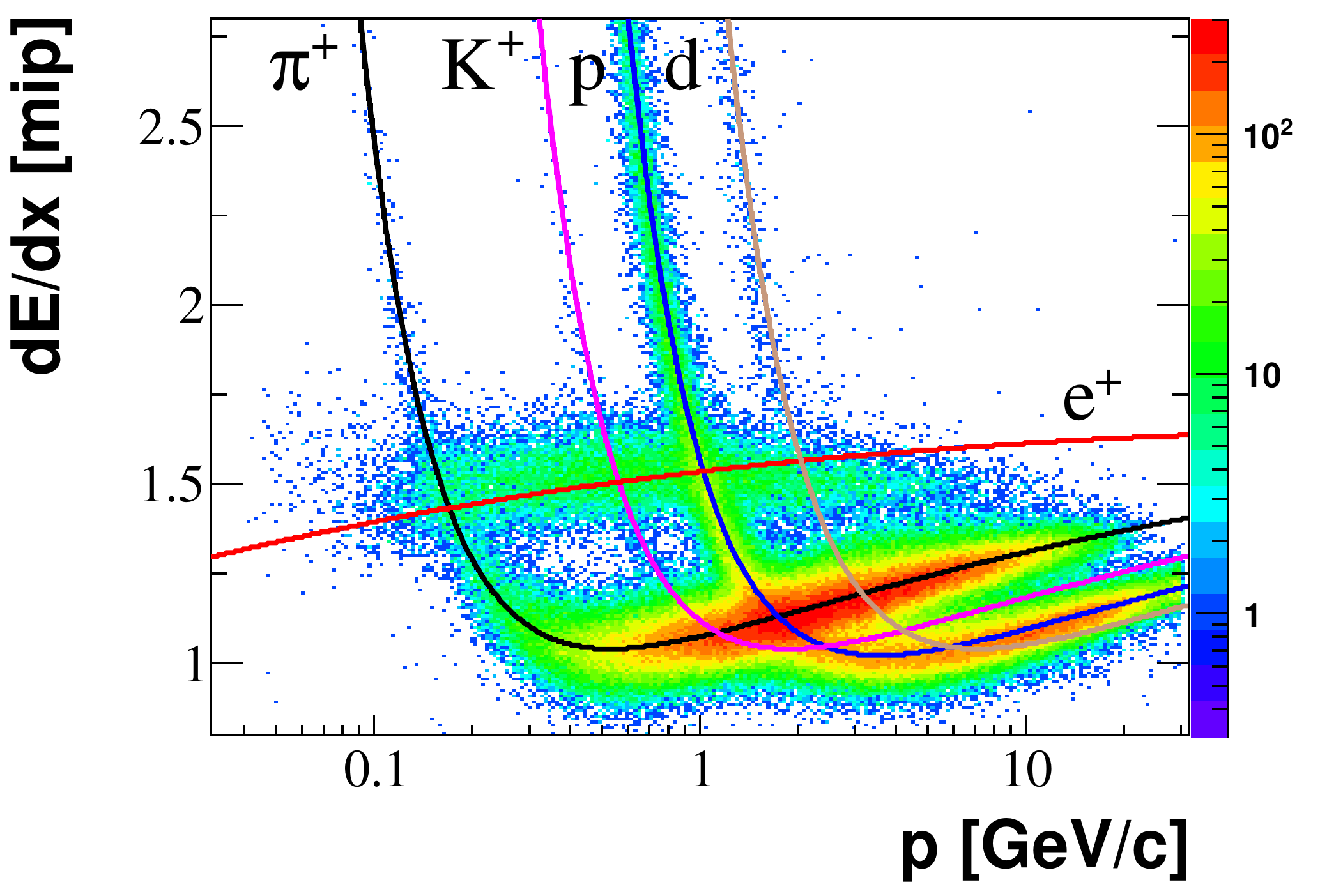}
\includegraphics[width=0.45\linewidth]{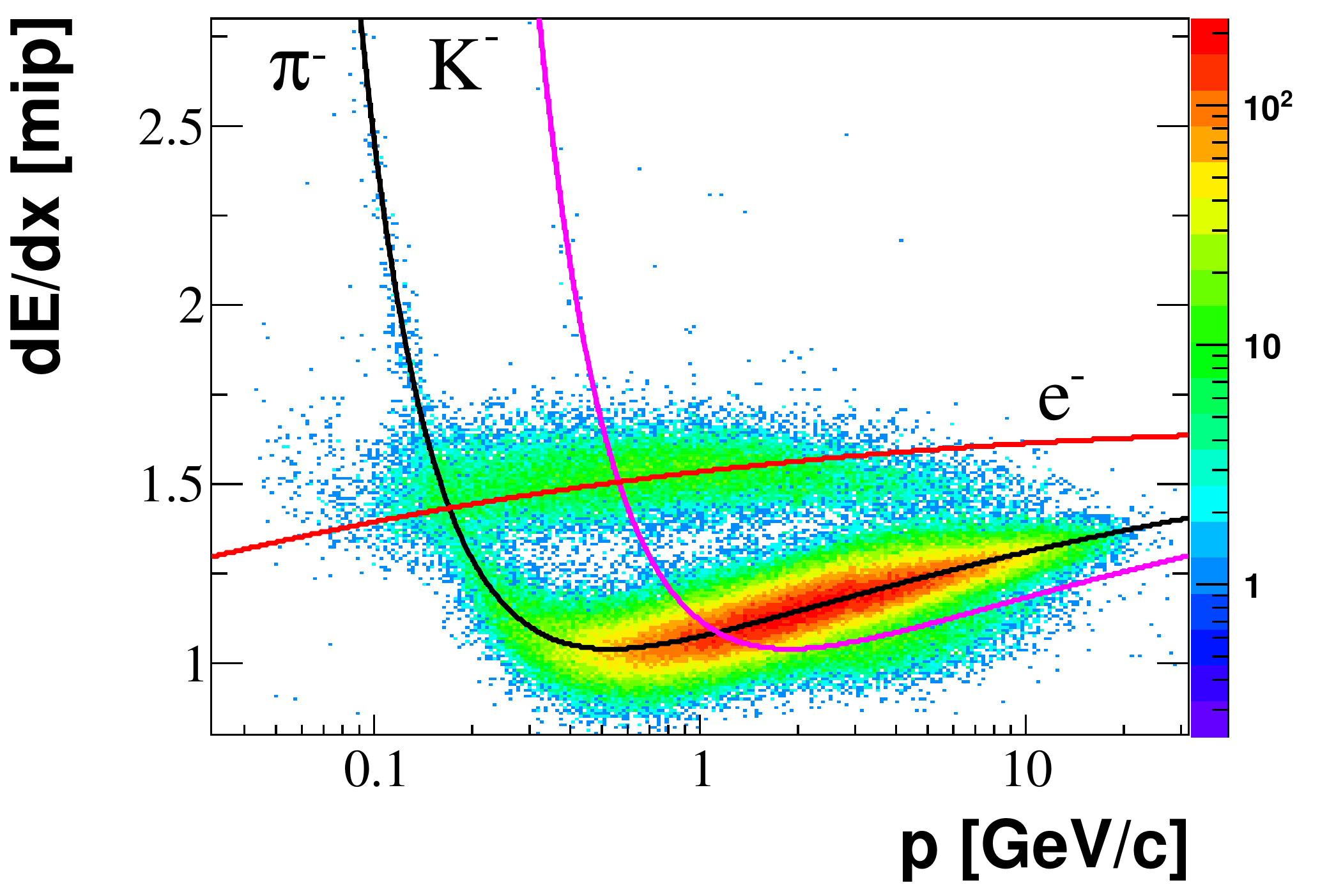}
\end{center}
\caption{(Color online) 
Specific energy loss in the TPCs for positively ($right$) and negatively ($left$) charged particles as a function of momentum. Curves show parameterizations of the mean $dE/dx$ calculated for different particle species.
}
\label{fig:dedx}
\end{figure}

This section presents the procedures used for data analysis to extract pion cross sections.
Crucial for this analysis is the identification of the produced pions.
Depending on the momentum interval, different approaches have been
adopted, which lead also to different track selection criteria.
The calibrated $dE/dx$ distributions as a function of particle momentum 
for positively and negatively
charged particles are presented in Fig.~\ref{fig:dedx}.

The task is facilitated for the negatively charged pions, by the
observation that more than 90\% of primary negatively charged particles
produced in p+C interactions at this energy are $\pi^-$, and thus the analysis
of $\pi^-$ spectra can also be carried out without additional particle
identification.

In the low momentum region (less than about 1~GeV/c), 
it is sufficient to distinguish pions
from electrons/positrons, kaons and protons by means of particle identification via
measurements of specific energy loss ($dE/dx$) in the TPCs.
A reliable  identification of $\pi^{+}$ mesons was not possible 
at momenta above 1~GeV/c where the Bethe-Bloch (BB) curves
for pions, kaons and protons cross each other. 
On the other hand, for $\pi^{-}$ mesons, where the contribution of $K^-$ and antiprotons 
is almost negligible, the $dE/dx$ analysis could be extended 
in momentum up to 3~GeV/c allowing consistency checks 
with the other analysis methods in the region of overlap.

\begin{figure}[h]
\begin{center}
\includegraphics[width=0.55\linewidth]{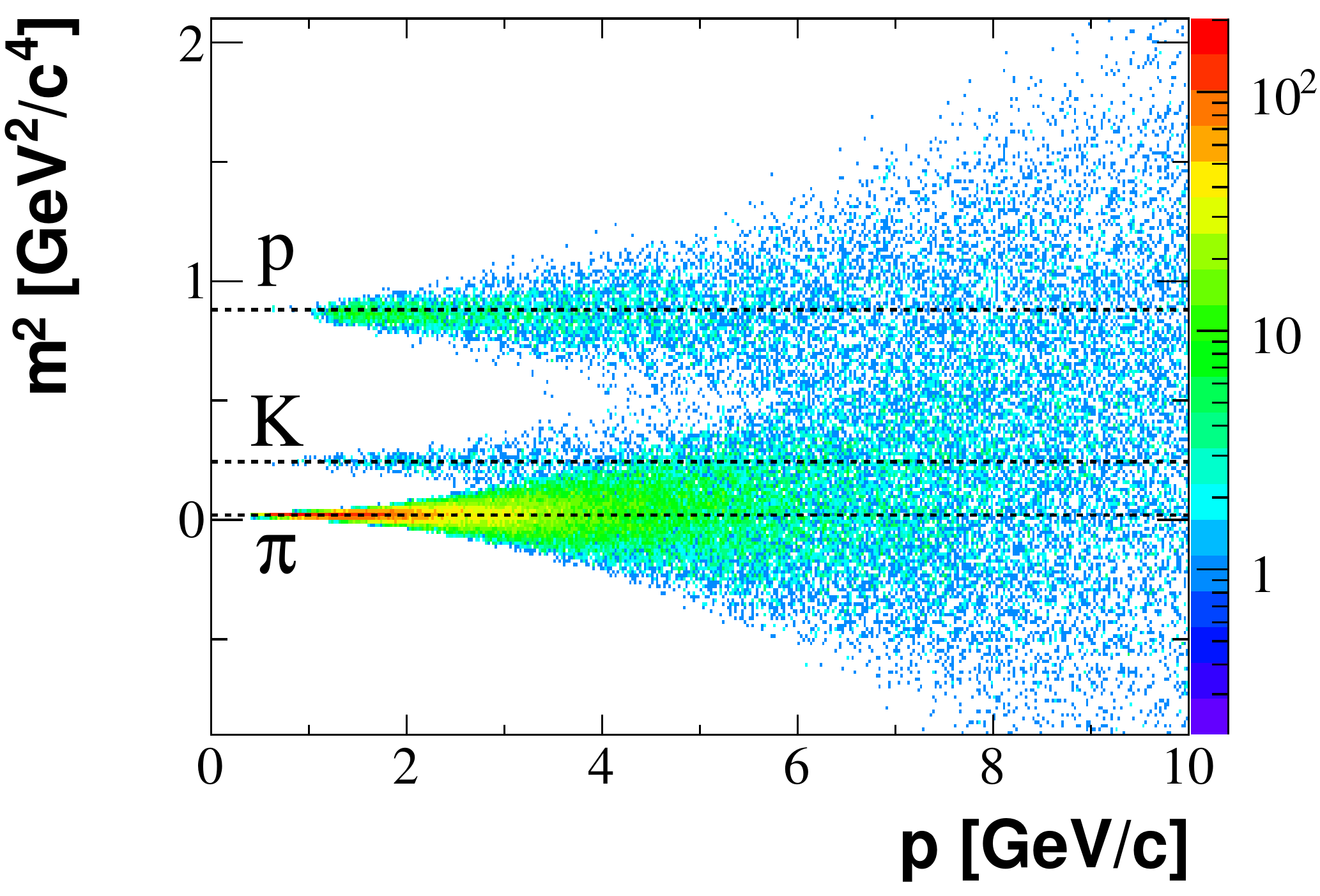}
\includegraphics[width=0.39\linewidth]{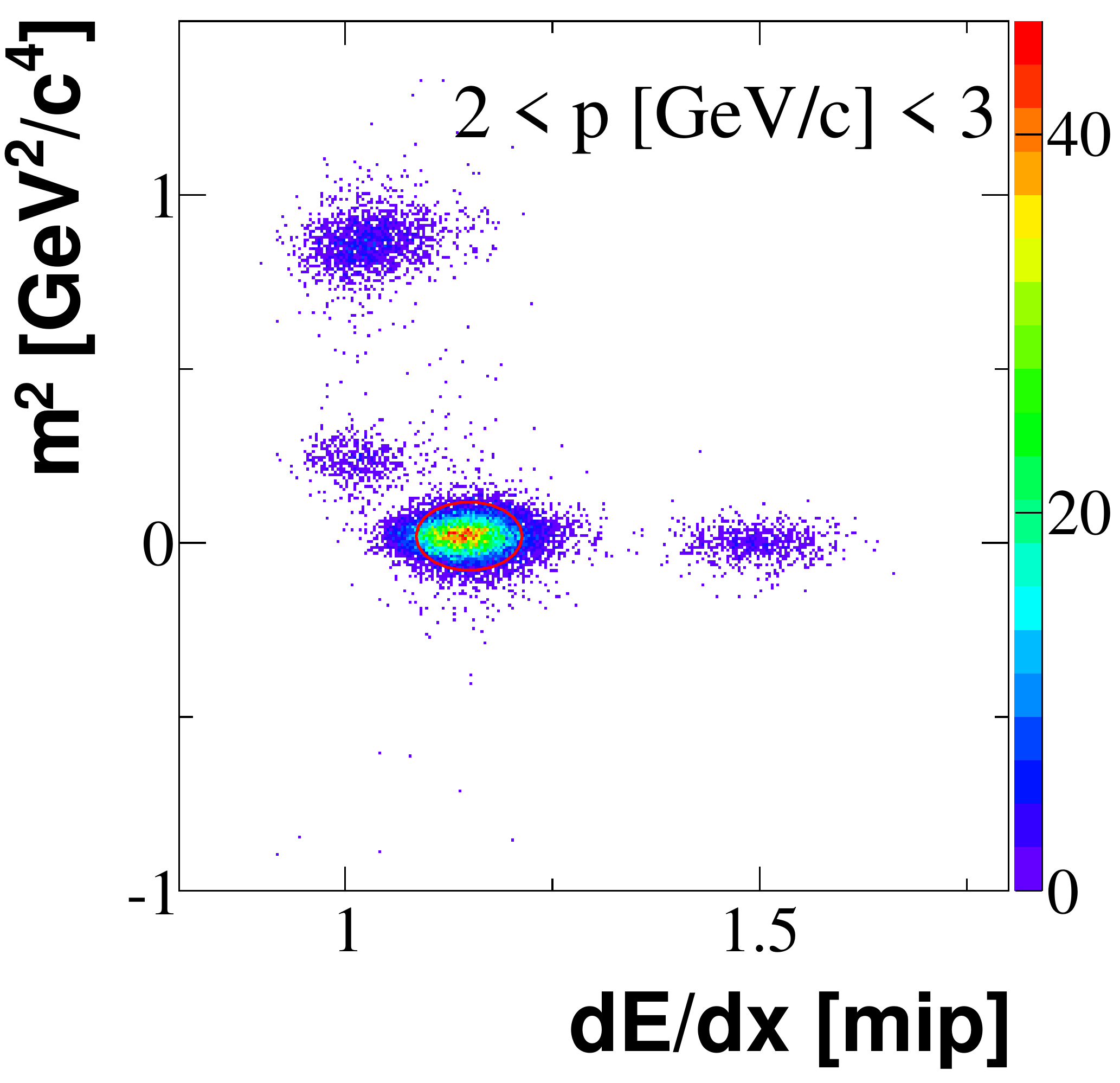}
\end{center}
  \caption{(Color online) 
   {\it Right:}
   Mass squared, derived from the ToF-F measurement and the fitted 
   path length and momentum, versus momentum $p$.
  The lines show the expected mass squared values for different particles.
  {\it Left:}
  Example of two-dimensional $m^2$--$dE/dx$ plots for
  positively charged particles in the momentum range 2-3~GeV/c. 
  $2\sigma$ contour around fitted pion peaks are shown.
}
\label{fig:ftof_resolution}
\end{figure}

High purity particle identification can be performed by combining
the $tof$ and $dE/dx$ information. Moreover, in the momentum range 
1-4~GeV/c, where $dE/dx$ bands for different particle species overlap,
particle identification is in general only possible using the $tof$ method, see
Figs.~\ref{fig:ftof_resolution}.
The ToF-F detector was designed to cover the necessary acceptance in
momentum and polar angle required by the T2K experiment, although limited to
particle momenta above about 0.8~GeV/c.

Indeed, three analysis methods were applied to obtain pion spectra:
\begin{enumerate}[(i)]
\setlength{\itemsep}{1pt}
\item analysis of $\pi^-$ mesons via measurements of
negatively charged particles ({\it $h^-$ analysis})~\cite{Tomek}
\item analysis of $\pi^+$ and $\pi^-$ mesons
identified via $dE/dx$ measurements
in the TPCs ({\it $dE/dx$ analysis at low momentum})~\cite{Magda} and
\item analysis of $\pi^+$ and $\pi^-$ mesons
identified via time-of-flight and $dE/dx$
measurements in the ToF-F and TPCs, respectively
({\it $tof-dE/dx$ analysis})~\cite{Sebastien}.
\end{enumerate}

Each analysis yields fully corrected pion spectra with
independently calculated statistical and systematic errors.
The spectra  were compared
in overlapping phase-space domains to check their consistency.
Complementary domains were combined to reach maximum acceptance.

The agreement between the spectra obtained by different methods 
is, in general, better than 10\%.
Note, that data points in the same ($p$,~$\theta$) bin
from different analysis methods
are statistically correlated as they result from the analysis of
the same data set.
In order to obtain the final spectra consisting of
statistically uncorrelated points the measurement with the
smallest total error was selected.

The analysis was based on a sample of 521~k events selected
from the total sample of 667~k of registered and reconstructed
events. This criterion essentially removes
a contamination by interactions upstream of the target.
For the event sample with the target removed, the selection reduces the
number of events from 46~k to 17~k.

\section{Results}
\label{Sec:results}

This section presents  results on inelastic and production
cross sections as well as on differential spectra of $\pi^+$ and $\pi^-$ mesons in
p+C interactions at 31~GeV/c.

\subsection{Inelastic and production cross sections}
\label{Sec:totalxsection}

The total inelastic cross section is measured to be~\cite{Claudia}
\begin{eqnarray*}
    \label{eq:finalsiginel}
\sigma_{inel} = 257.2 \pm 1.9 \pm 8.9~\mathrm{mb}~.
\end{eqnarray*}

The production cross section was calculated from the inelastic
cross section by subtracting the quasi-elastic contribution~\cite{ref:Glauber}.
The result is~\cite{Claudia}:
\begin{eqnarray*}
    \label{eq:finalsigprod}
\sigma_{prod} = 229.3 \pm 1.9 \pm 9.0~\mathrm{mb}~.
\end{eqnarray*}

\begin{figure}[!h]
\begin{center}
\includegraphics[width=0.7\linewidth]{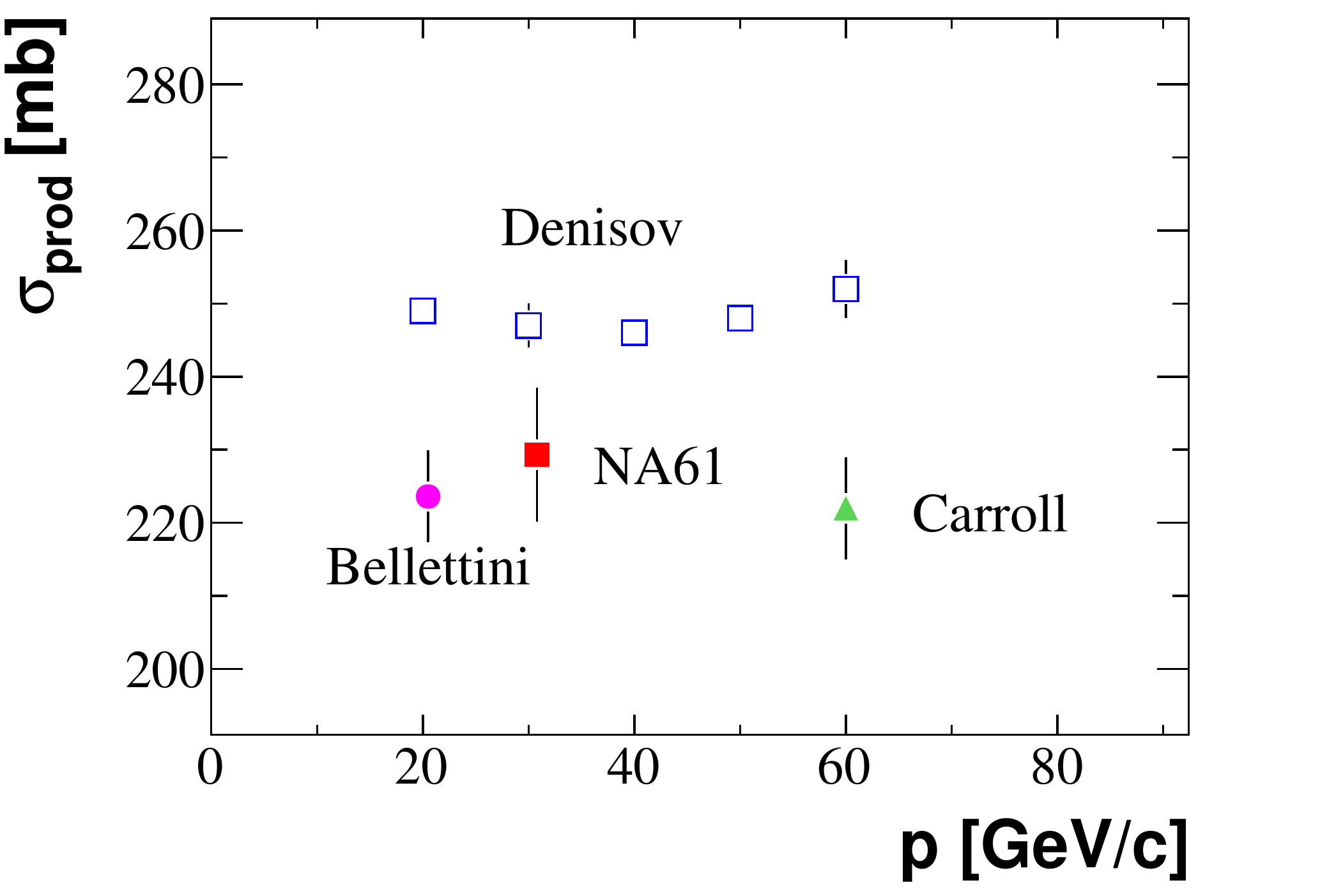}
\end{center}
 \caption{(Color online) 
 Beam momentum dependence of the production cross section
 for p+C interactions. The NA61/SHINE result (filled square)
 is compared with previous measurements:
 Bellettini et al. (circle)~\cite{ref:Bellettini},
 Carroll et al. (triangle)~\cite{ref:Carroll} and Denisov et al. (open squares)~\cite{ref:Denisov}.
 For the NA61/SHINE point, the error bar indicates statistical and systematic uncertainties
  added in quadrature. The result from Ref.~\cite{ref:Bellettini} was recalculated by
 subtracting from the measured inelastic cross section
 a quasi-elastic contribution at 20~GeV/c of ($30.4 \pm 1.9~(sys)$)~mb.
}

\label{fig:inelastic}
\end{figure}

The production cross section is compared to previous measurements
in Fig.~\ref{fig:inelastic}. 

\subsection{Spectra of $\pi^+$ and $\pi^-$ mesons}

\begin{figure*}[tb]
\begin{center}
\includegraphics[width=0.9\linewidth]{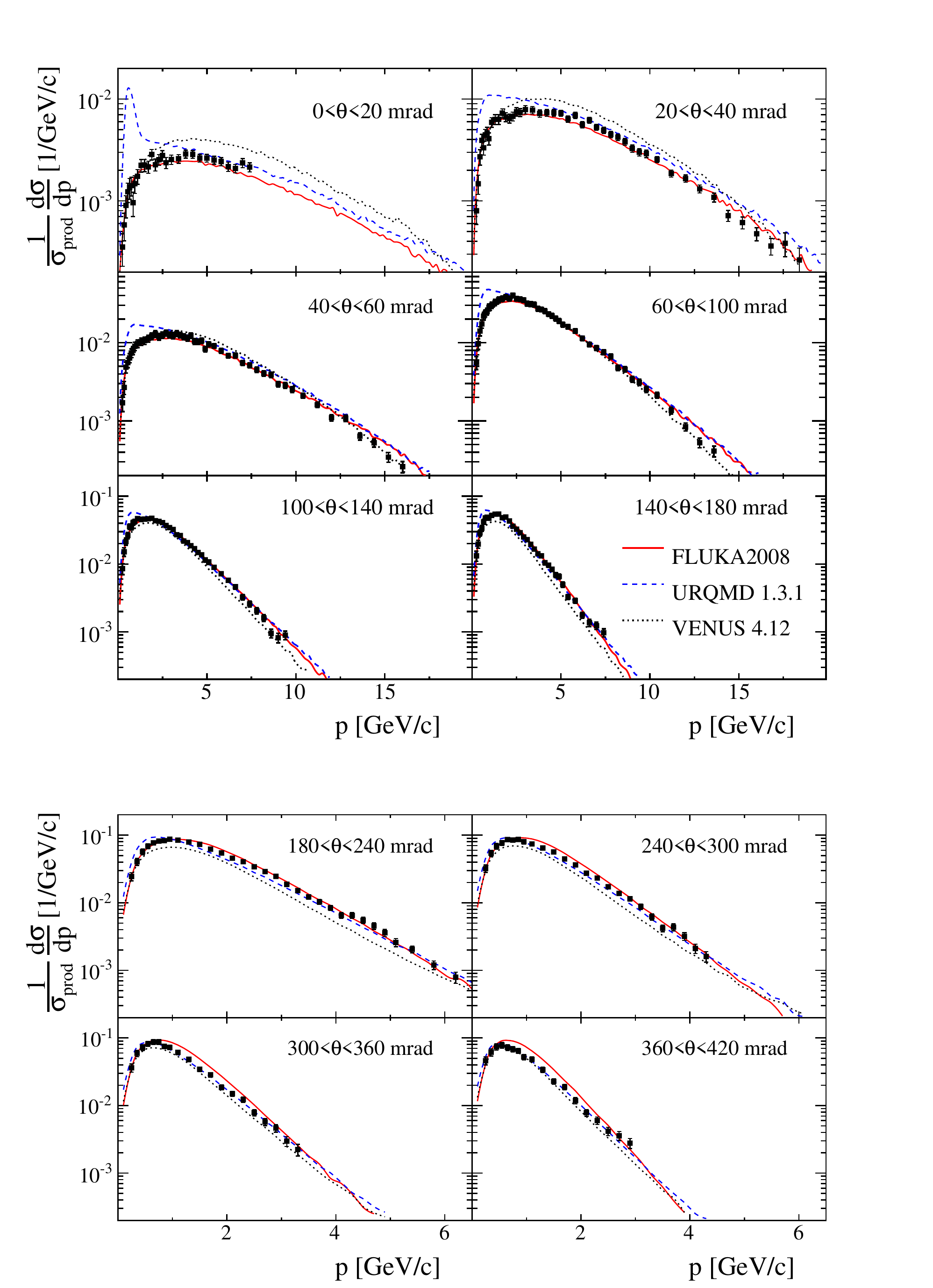}
\caption{(Color online) 
  Laboratory momentum distributions of $\pi^{+}$ mesons produced
  in production p+C interactions at 31~GeV/c
  in different intervals of polar angle ($\theta$).
  The spectra are normalized to the mean $\pi^{+}$ multiplicity in
  all production p+C interactions.
  Error bars indicate statistical and systematic uncertainties
  added in quadrature.
  The overall uncertainty~($2.3\%$) due to the normalization procedure is not shown.
  Predictions of hadron production models,
  \FlukaLong~\cite{Fluka} (solid line),
  \UrqmdLong~\cite{Urqmd} (dashed line) and \VenusLong~\cite{Venus} (dotted line)
   are also indicated.
}
\label{pion_minus_model}
\end{center}
\end{figure*}

\begin{figure*}[tb]
\begin{center}
\includegraphics[width=0.9\linewidth]{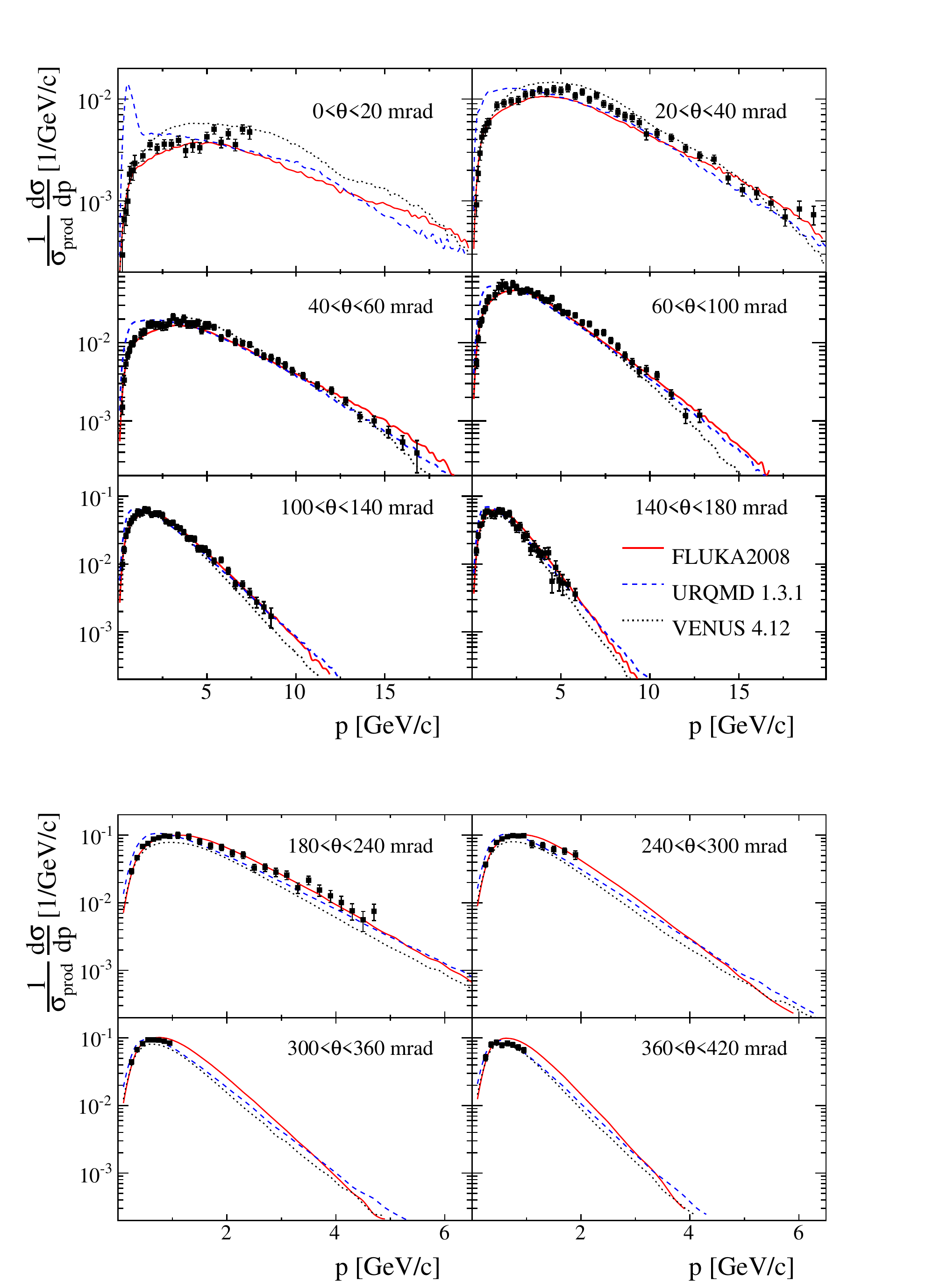}
\caption{(Color online) 
  Laboratory momentum distributions of $\pi^{-}$ mesons produced
  in production p+C interactions at 31~GeV/c
  in different intervals of polar angle ($\theta$).
  The spectra are normalized to the mean $\pi^{-}$ multiplicity in
  all production p+C interactions.
  Error bars indicate statistical and systematic uncertainties
  added in quadrature. 
  The overall uncertainty~($2.3\%$) due to the normalization procedure is not shown.
  Predictions of hadron production models,
  \FlukaLong~\cite{Fluka} (solid line),
  \UrqmdLong~\cite{Urqmd} (dashed line) and \VenusLong~\cite{Venus} (dotted line)
   are also indicated.
}
\label{pion_plus_model}
\end{center}
\end{figure*}

The $\pi^+$ and $\pi^-$  spectra presented in this
section refer to pions produced  in strong and
electromagnetic processes in
p+C interactions at 31~GeV/c.

The spectra are  presented as a function of particle momentum
in ten intervals of the polar angle.
Both quantities are calculated in the laboratory system.
The chosen binning takes into account the available statistics
of the  2007 data sample, detector acceptance and particle production kinematics.

The final spectra are plotted in Figs.~\ref{pion_minus_model} and ~\ref{pion_plus_model}.
For the purpose of a comparison of the data with model
predictions the spectra were normalized to the
mean $\pi^\pm$ multiplicity in all
production interactions by dividing by $\sigma_{prod}$.
This avoids uncertainties due
to the different treatment of quasi-elastic interactions
in models as well as problems due to the absence of predictions for
inclusive cross sections.

\section{Conclusion}

This work presents inelastic and production
cross sections as well as positively and negatively
charged pion spectra in p+C interactions at 31~GeV/c on the 2007 thin (4\% interaction length) carbon target.
These data are essential for  precise predictions of the 
neutrino flux for the T2K long baseline neutrino oscillation experiment in Japan.
Furthermore, they provide important input to improve hadron production models
needed for the interpretation of air showers initiated by ultra high
energy cosmic particles.

A much larger data set with both the thin and the T2K replica carbon
targets was recorded in 2009 and 2010 and is presently being analysed.
This will lead
to results of higher precision for pions and extend the measurements to
other hadron species such as charged kaons, protons, $K^0_S$ and $\Lambda$. 
Analysis of the data collected with the T2K replica target during the 2007 run is in progress~\cite{Nicolas}.
The new data will allow a further significant reduction of the uncertainties
in the prediction of the neutrino flux in the T2K experiment.

\section{Acknowledgments}
This work was supported by the following funding agencies:
the Hungarian Scientific Research Fund (grants OTKA 68506 and 79840),
the Polish Ministry of Science and Higher Education (grants 667/N-CERN/2010/0,
N N202 1267 36, N N202 287838 (PBP 2878/B/H03/2010/38), DWM/57/T2K/2007),
the Federal Agency of Education of the Ministry of Education and Science
of the Russian Federation (grant RNP 2.2.2.2.1547), 
the Russian Academy of Sciences and
the Russian Foundation for Basic Research (grants 08-02-00018 and 09-02-00664),
the Ministry of Education, Culture, Sports, Science and Technology,
Japan, Grant-in-Aid for Scientific Research (grants 18071005, 19034011,
19740162, 20740160 and 20039012), the Toshiko Yuasa Lab. 
(France-Japan Particle Physics Laboratory),
the Institut National de Physique Nucl\'eaire et Physique des Particules
(IN2P3, France),
the German Research Foundation (grant GA 1480/2-1),
the Swiss National Science Foundation 
(Investigator-Driven projects and SINERGIA) and the Swiss State Secretariat 
for Education and Research (FORCE grants). \\
The authors also wish to acknowledge the support provided 
by the collaborating institutions, in particular,
the ETH Zurich (Research Grant TH-01 07-3), 
the University of Bern and the University of Geneva.

Finally, it is a pleasure to thank 
the European Organization for Nuclear Research 
for a strong support and hospitality
and, in particular, the operating crews of the CERN SPS accelerator
and beam lines who made the measurements possible.  

\bibliographystyle{ws-procs975x65}

\end{document}